\documentstyle[epsf,12pt]{article}
\begin{document}

\setlength{\baselineskip}{0.30in}
\def\psl{p \hspace*{-0.5em}/}
\def\ksl{k \hspace*{-0.5em}/}
\newcommand{\be}{\begin{equation}}
\newcommand{\bi}{\bibitem}
\newcommand{\al}{\alpha_{el}}
\newcommand{\ea}{\alpha_{el}}
\newcommand{\am}{\alpha (\mu^2)}
\newcommand{\as}{\alpha_s}
\newcommand{\ak}{\alpha_s(k^2)}
\newcommand{\ap}{\alpha_s (k_{\perp}^2)}
\newcommand{\apz}{\alpha_s ({k_{\perp}^2\over 1-z})}
\newcommand{\bb}{\beta}
\newcommand{\bef}{\beta -{\rm function}}
\newcommand{\kp}{k_{\perp}}
\newcommand{\de}{\delta}
\newcommand{\La}{\Lambda_{QCD}}
\newcommand{\LU}{\Lambda_{UV}}
\newcommand{\aL}{\alpha(\La )}
\newcommand{\ee}{\end{equation}}
\newcommand{\aQ}{\alpha_s(Q^2)}
\newcommand{\Nc}{N_{crit}}
\newcommand{\De}{\Delta}
\newcommand{\om}{\omega}
\newcommand{\vp}{\varphi}
\newcommand{\la}{\lambda}
\newcommand{\si}{\sigma}
\newcommand{\eps}{\epsilon}
\newcommand{\bea}{\begin{eqnarray}}
\newcommand{\eea}{\end{eqnarray}}
\newcommand{\gc}{\langle 0|\as(G^a_{\mu\nu})^2|0\rangle}
\newcommand{\ra}{\rightarrow}
\newcommand{\ot}{\beta (\alpha )\approx b_0\alpha^2}
\newcommand{\qb}{\bar{q}}
 %SLASHCHAR
\def\slashchar#1{\setbox0=\hbox{$#1$}           % set a box for #1
   \dimen0=\wd0                                 % and get its size 
   \setbox1=\hbox{/} \dimen1=\wd1               % get size of /
   \ifdim\dimen0>\dimen1                        % #1 is bigger 
      \rlap{\hbox to \dimen0{\hfil/\hfil}}      % so center / in box 
      #1                                        % and print #1
   \else                                        % / is bigger 
      \rlap{\hbox to \dimen1{\hfil$#1$\hfil}}   % so center #1
      /                                         % and print /
   \fi}                                         %
%%EXAMPLE    $\slashchar{E}_{t}$ will put slash across the E only

\begin{flushright}
UM-TH-97-05\\
hep-ph/9701378\\
 \today
\end{flushright}

\begin{center}
\vglue .06in
{\Large \bf {Renormalon Variety in Deep Inelastic Scattering.}}\\ [.5in]

{\bf R. Akhoury and V.I. Zakharov}\\
[.15in]

{\it The Randall Laboratory of Physics\\
University of Michigan\\
Ann Arbor, MI 48109-1120}\\
[.15in]

\end{center}
\begin{abstract}
\begin{quotation}
We discuss the renormalon-based approach to power corrections
in non-singlet deep inelastic scattering structure functions
and compare it with the general
operator product expansion. The renormalon technique and
its variations  
relate the power corrections directly to 
infrared-sensitive parameters such as the position of the Landau pole 
$\La$ or the infinitesimal gluon mass $\la$. 
In terms of the standard OPE these techniques unify evaluations of the
coefficient functions and of matrix elements. We argue
that in case of deep inelastic scattering there is a proliferation
of competeing infrared sensitive parameters. In particular we consider
the gluon and quark masses, virtuality of quarks and  $\La$
as possible infrared cut offs and compare the emerging results.
In the standard renormalon technique where $\La$ is the infrared
parameter, the argument of the running coupling is crucial to 
obtain the correct $x$ dependance of the structure functions.
 Finally we discuss the limitations
of the use of the renormalon based methods  for determining
of the $x$ dependance of the power corrections. 

\end{quotation}
\end{abstract}
\newpage

\section{Introduction }

Generically, renormalons 
allow for a parametrization of infrared sensitive contributions
to hard processes in QCD within the framework of
perturbation theory (for the basic ideas see \cite{mueller,vz,azr}  and references therein ).     
If one considers only infrared safe observables then the infrared sensitive contributions are
power suppressed. Moreover, in many cases it does not matter which particular infrared
parameter is used in calculations. It can be, for example, the position
of the Landau pole in the  running coupling, $\La$, or (at the one-loop level) 
a fictitious gluon mass $\la$.
Since  renormalons are a pure perturbative construct the answer
is obtained directly in terms of the  infrared parameters chosen.
If, on the other hand, the treatment of the same process is possible
within an operator product expansion the evaluation of the power
corrections is a two-stage procedure. First, one calculates perturbatively
a coefficient function in front of operators. The matrix elements of these
operators provide then a measure of the infrared contribution,
both perturbative and non-perturbative.
Thus, renormalons fix the matrix elements
perturbatively. Finally, one adjusts the overall scale
of the matrix element to allow for its non-perturbative
enhancement. This tacit assumption is behind
almost all the applications of renormalons \cite{az2}. 

The best known example of this kind is the gluon condensate
$\gc$. It can be treated non-perturbatively \cite{svz}.
On the other hand if one introduces a finite gluon mass $\la$
then \cite{chet}:
\be
\gc_{\la \neq 0}~=~-{3\as \over \pi^2}\la^4 ln\la^2\label{ch}
\ee
where we keep only the term nonanalytic in $\la^2$
since only such terms can be consistently attributed to 
the infrared region. 
Instead of introducing $\la\neq 0$ one can evaluate 
the gluon condensate associated with a renormalon chain \cite{mueller,vz}:
\be
\gc_{ren}~=~p.v. \int{3d^4k\over \pi^2}\as (k^2)~=~const\cdot\La^4\label{rc}\label{lp}
\ee
where $\ak$ is the running coupling corresponding to a single-term
$\beta$-function and the integration 
over the pole in $\as (k^2)$  has been defined as the principal value (p.v.).
In this example the use of renormalons, i.e. of Eqs. (\ref{ch}),
(\ref{rc}), does not enhance the predictive power of the theory.
Indeed, Eqs. (\ref{ch}), (\ref{rc}) cannot be taken literally
and one reserves for an unknown rescaling of the matrix element
to fit the data. What renormalons do achieve is a short cut of the procedure since, say, terms
of order $\la^4ln\la^2$ can be evaluated directly for
physical quantities without invoking the operator product expansion as an intermediate step. 
This may be a decisive advantage of renormalons in cases where there is
no OPE available. 

Thus at first sight the alternative is that either there is an 
OPE and then the renormalons are a particular model for 
the matrix elements involved or there is no OPE and then the renormalons
could be a substitution for it. 
This logic appears to be defied by the example
of deep inelastic scattering. Indeed, 
the power corrections can be treated either via  the OPE \cite{vs,soldate,ellis}
and or via renormalons \cite{mueller1,dmw,stein,dasgupta}.
While the OPE reserves for at least one unknown matrix element for each
moment, renormalons are claimed to fix the whole $x$ dependence of the
$1/Q^2$ correction \cite{dmw,stein,dasgupta}.
Moreover the same approach can be used to study the power corrections to fragmentation
functions \cite{bb}.

 Motivated by these very interesting observations we look, in this paper
\cite{footnote},  into the anatomy of the infrared sensitive contributions 
in the renormalon calculus as applied to DIS.
We find that already at the one-loop level there are a variety
of possible infrared procedures which do not obviously give
the same answer.  For example,  as discussed in more detail later in the paper, 
one possible strategy 
underlying the applications of renormalons is to reduce the matrix elements
governing higher twist contributions to those of the leading twist.
Typically, the reduction produces a factor $f$ proportional to:
$$
f(\eps^2,m^2,\la^2)
~\sim~\int_0^1dyX(y)ln(X(y))
$$ where
$$
X(y)~=~{\eps^2\over Q^2}y(y-1)+{m^2\over Q^2}y^2+{\la^2\over Q^2}(1-y)
$$
$\la,m$ are the gluon and quark masses respectively and
$\eps$ is the virtuality of the quark, $p^2-m^2\equiv\eps^2$. 
This proliferation of the infrared sensitive parameters, that is  $\la,m,\eps$,
is due to the fact that there are both soft gluon and soft
quark lines. On the other hand the classical 
applications of renormalons assume that there is only a single gluon
line made soft through an insertion of a renormalon chain, while the other lines are hard. 
Thus in case of DIS, the renormalon-related parameter $\la$ determines the result 
only upon forcing the other parameters ($m,\eps$) to vanish -
$m^2,\eps^2\ll \la^2$,  which is not necessarily natural.
One could of course assume, for example, that to the contrary,  $m^2\gg \la^2$.
In addition to $\la$ and $m$, one could also explore the infrared sensitivity by the standard
renormalon chain technique in which the power corrections are proportional to ${\La}^2/Q^2$,
etc.     In sections 2 , 3 and 4 we consider in more detail the $x$ dependance of the non-singlet
structure functions with $\la$, $\La$ and $m$ , respectively, as dominating infrared
parameters.   Section 2 essentially reproduces the results of \cite{dmw,stein,dasgupta}. In
section 3 we find that not only is the predicted $x$  dependance a function of what the
argument of the running is, but in addition, two arbitrary scales typified by two
different integrals multiplying ${\La}^2/Q^2$ are introduced, in general.  This
difference in fact arises because some contributions come from Feynman integrals which are 
collinear divergent whereas others from ones that are not.  This is to be
contrasted with the case discussed in section 2 where there is just one unknown
parameter $\la$ and is an indication of an inherent infrared instability.                        
 As for the dependence on the virtuality $\eps$ it  also indicates that
in fact we are dealing with an infrared unsafe quantity. 
This would, in general, become manifest in the two-loop approximation
for the power corrections. To see this one does not need to evaluate the second loop explicitly
but it is enough to note that the anomalous dimensions of the operators
governing the leading and higher twist contributions 
are different. This is discussed in section 4 . In section 5 we present our conclusions.

\section{Gluon Mass as an Infrared Cut- Off.}

As is mentioned in the introduction there exist a variety of choices
of infrared sensitive parameters. 
Thus far the dispersive approach to the running coupling \cite{dmw}, and 
the renormalon chain in the large $N_f$ limit \cite{stein} have been tried ,
with identical conclusions for the $x$ dependance of the DIS structure functions wherever
there is overlap. 
To these we will add the cases of
$\la,m\neq 0$ and of the Landau pole contribution in the running coupling.
These cases are considered in this and subsequent sections.

To introduce notations, we consider the partonic structure tensor defined as
\be
W_{\mu\nu}(p,q)~=~{1\over 8\pi}\sum M_{\mu}M^{*}_{\nu}\ee
where $M_{\mu}$ is the amplitude for $\gamma^{*}+q\ra q'+g+...$
where $q$ is the parton ( quark for our purposes ) of momentum $p$. An average over the
initial quark spins is understood and the standard decomposition of the $W_{\mu\nu}$ is used:
\bea
W_{\mu\nu}~=~{1\over 2}\left(g_{\mu\nu}-{q_{\mu}q_{\nu}\over q^2}\right) F_L(z,Q^2)~+\\
\nonumber
\left(p_{\mu}p_{\nu}-{p\cdot q\over q^2}(p_{\mu}q_{\nu}+p_{\nu}q_{\mu})
+g_{\mu\nu}{(p\cdot q)^2\over q^2}\right){F_2(z,Q^2)\over p\cdot q}
\eea
where $z=Q^2/2p\cdot q,~ q^2=-Q^2$ and $p^2=0$.
Note that we put the mass of the quark exactly zero while keeping a finite gluon mass
so as to ensure the role of $\la\neq 0$ as a unique infrared cut off.
For the structure function $F_L$ only the diagram
with emission of a real gluon contributes to lowest order in $\as$.
For single gluon emission we have :
\be
W_{\mu\nu}={C_F\as\over2}\int{d^4k\over(2\pi)^4}
(2\pi)^2\de_+((k+q)^2)\de_+((p-k)^2-\la^2)
{d_{\rho\si}(p-k)\over k^4}
Tr\left[\slashchar{p}\gamma_{\rho}\slashchar{k}\gamma_{\mu}(\slashchar{k}+\slashchar{q})
\gamma_{\nu}
\slashchar{k}\gamma_{\si}\right]
\ee
where in the Feynman gauge $d_{\rho\si}=-g_{\rho\si}$.

To perform the integral we use a Sudakov parametrization:
\be
k^{\mu}~=~\kappa p^{\mu}+{k^2+k_{\perp}^2\over 2\kappa}n^{\mu}+k_{\perp}^{\mu}\label{kap}
\ee
where $n^2=0,p^2=0,n\cdot k_{\perp}=0,n\cdot p=1,p\cdot k_{\perp}=0$  and  we may choose
\be
n^{\mu}~=~{q^{\mu}+zp^{\mu}\over p\cdot q}.
\ee
Moreover, for these variables:
\be
\int d^4k~=~{\pi\over 2}\int{d\kappa\over\kappa}dk^2dk_{\perp}^2\label{ps}.
\ee
For detecting the non-analytical terms the limits of the $k^2$ integration are also
important:
\be
{\la^2z\over 1-z}~\le~k^2~\le~{Q^2\over z}-\la^2.\label{limits}
\ee

It is straightforward to find for $F_L$:
\be
F_L~=~2C_F{\as\over 4\pi}
{4z^2\over Q^2}\int {dk^2\over k^4}(k^2-\la^2)^2\left(1+{k^2-\la^2\over Q^2}z\right),
 \label{fl1} 
\ee
with the limits of integration as indicated in (\ref{limits}).
From (\ref{fl1}) we immediately get for the term independent of $\la$ and
representing therefore the leading twist:
\be
F_L~=~C_F{\as\over 4\pi}4z
\ee
which is a well known result.
Furthermore for the terms $\la^2ln\la^2$ and $\la^4ln\la^2$ we get:
\be
(F_L)_{\la^2}
~=~C_F{\as\over 2\pi}{4z^2\over Q^2}\cdot 2\la^2ln\la^2\label{ls}
\ee
and
\be
(F_L)_{\la^4}~=~C_F{\as\over 2\pi}{4z^3\over Q^4}(-3)\la^4ln\la^2\label{lf},
\ee
respectively.
It is worth emphasizing again that equations (\ref{ls},\ref{lf}) are
understood in the sense that only non-analytic terms in $\la^2$ are kept.

The results (\ref{ls}) and (\ref{lf}) do reproduce the predictions 
based on the dispersive approach to the running coupling \cite{dmw}.
This coincidence of the results comes as no surprise since the dispersive formulation
is, in fact, aimed  at improving the high-energy behaviour of the theory
with massive gluons. However, the change in this high-energy behaviour obviously
does not affect the infrared sensitive pieces which are non-analytic in $\la^2$.

\section{Landau-Pole Parametrization.}

As the next way to parametrize the infrared sensitive contributions
we examine in this section the contribution of the Landau pole
in the running coupling,  in which case the power corrections are proportional to
$({\La}^2 /Q^2)^n$.
For definiteness we again consider first the longitudinal structure function.
Since now the infrared sensitive parameter will be $\La$ we can put $\la=0$. 
Moreover, we  now account for the running of the effective coupling
while still using the kinematics of one-gluon emission.
The crucial issue then is what is the argument of the running coupling.
We perform first the integration assuming that it is $k_{\perp}^2$
that determines the running, i.e, $\ap$. We shall
see however that to reproduce the results of the previous sections
one should assume that the true argument of the effective coupling
is $k^2_{\perp}/(1-z)$.

To depict the contribution to $F_L$ of the Landau pole we first
integrate over $\kappa$  and $k^2$ (see Eqs. (\ref{kap}), (\ref{ps})).
Thus we start with
\bea
F_L~=~2C_F {4z^2\over Q^2}\int~{\ap\over 4\pi}
{d\kp^2}{d\kappa\over \kappa}{dk^2\over k^4}
\de\left({\kp^2\over\kappa}+{k^2(1-z)\over \kappa}\right)\cdot\\ \nonumber
\de\left(k^2+{\kappa\over z}Q^2-Q^2-{k^2+\kp^2\over\kappa}z\right)
\left({k^2+\kp^2\over \kappa}\right)^2
\left({k^2+\kp^2\over \kappa}+{Q^2\over z}\right).
\eea
Then the integration over $k^2$ is trivial while the integration over $\kappa$
can be expressed in terms of $\kappa_{\pm}$:
\be
2\kappa_{\pm}~=~
1+z\pm\sqrt{(1-z)^2-4(\kp^2/Q^2)z(1-z)}.
\ee
In this way we come to the following integral over $\kp^2$:
\be
F_L~=~2C_F\cdot {4z^2\over Q^2}\int_0^{Q^2(1-z)/4z}
{\ap \over 4\pi}
{d\kp^2\over \sqrt{(1-z)^2-4(\kp^2/Q^2)\label{renorm}
z(1-z)}}.
\ee

In order to pick up the renormalon contributions we 
substitute the following representation  in  Eq. (\ref{renorm}) :
\be
\ap~=~\int_0^{\infty}d\si\left({\kp^2\over \La^2}\right)^{-\si \beta_1}\label{borel}
\ee
where $\beta_1$ is the first coefficient in the $\bef$.
Furthermore, to perform the integration over $\kp$ it is convenient to
introduce a new variable, $y=4z(1-z)^{-1}\kp^2Q^{-2}$
which allows us to disentangle the $z$ and $\kp$ dependences:
\be
F_L~=~{2C_F\over 4\pi}
\cdot {4z^2\over 1-z}\int_0^{\infty}d\si\left({\La^2\over Q^2}\right)^{\si \beta_1}
\left({1-z\over 4z}\right)^{1-\si \beta_1}\int 
dy~y^{-\si \beta_1}(1-y)^{-1/2}.
\ee
Finally we get for $F_L$ the following expression
\be
F_L~=~{2C_F\over 4\pi}
\cdot {4z^2\over 1-z}\int_0^{\infty}d\si\left({\La^2\over Q^2}\right)^{\si \beta_1}
\left({1-z\over 4z}\right)^{1-\si \beta_1}
{\Gamma(1-\si \beta_1)\Gamma (1/2)\over \Gamma(3/2-\si \beta_1)} \label{fl}
\ee
which exibits the infrared renormalons corresponding to the poles of 
$\Gamma (1-\si \beta_1)$. Since 
$(\La^2/Q^2)^{\sigma \beta_1}\sim exp(-\si/\as(Q^2/\La^2))$ the corresponding power
ambiguities are proportional to $(\La^2/Q^2),(\La^2/Q^2)^2...$

In order to extract the power behaviour one must define the integral over $\sigma$ using some
prescription.  We next outline how this  integral  may be defined via, say,  a principal
value prescription for the poles at $\sigma=n/\beta_1$, $n=1,2,...$. First we note that small
values of $\sigma$ correspond to large values of $\kp$ and hence for this we get the usual
renormalization group improved perturbative answer . Keeping this in mind
we may divide the integration region thus:
\begin{equation}
\sigma~\in~[0,\infty]~=~[0, {s \over \beta_1}] + \sum_{n \geq 1}[{{n-s} \over \beta_1},
{{n+s} \over \beta_1}],
\end{equation}
where, $0<s<1$, and a typical choice could be $s=1/2$.
The first region gives the perturbative answer, as we have checked, and the rest give the power
corrections in the principal value prescription. To make this explicit, consider the contribution
from the pole at $\sigma = n/\beta_1$ to the integral (see Eq. (\ref{fl})):
\begin{equation}
\int_0^{\infty}d\si\left({4z\La^2\over (1-z)Q^2 }\right)^{\si \beta_1}
{\Gamma(1-\si \beta_1)\Gamma (1/2)\over \Gamma(3/2-\si \beta_1)}
\end{equation}
which is defined by:
\begin{equation}
\int_{{n-s} \over \beta_1}^{{n+s} \over \beta_1}d\si\left({4z\La^2\over (1-z)Q^2
}\right)^{\si\beta_1} 
{\Gamma(1-\si \beta_1)\Gamma (1/2)\over \Gamma(3/2-\si \beta_1)}
\end{equation}
This can be written after some algebra as:
\begin{equation}
\left({4z\La^2\over (1-z)Q^2}\right)^n  {\bar{I}}_n^{pv}
\end{equation}
where, the integral ${\bar{I}}_n^{pv}$ is defined by :
\begin{equation}
{\bar{I}}_n^{pv}= (-1)^n{{\pi}^{3/2} \over \beta_1}\int_0^{s}{d\sigma \over
sin\pi\sigma}\left[{(4z{{\La}^2 \over (1-z)Q^2})^{\sigma}(n+\sigma) \over
{\Gamma(1+n+\sigma)\Gamma(3/2-n-\sigma)}}-
{(4z{{\La}^2 \over (1-z)Q^2})^{-\sigma}(n-\sigma) \over
{\Gamma(1+n-\sigma)\Gamma(3/2-n+\sigma)}}\right] 
\end{equation}

 From this, for the first two leading power corrections we finally get:
\begin{eqnarray}
(F_L)_{1/Q^2}~=~{C_F \over 2\pi}{4z^2\over 1-z}{\La^2\over Q^2}{\bar{I}}_1^{pv}, \nonumber \\
(F_L)_{1/Q^4}~=~{C_F \over 2\pi}{16z^3\over (1-z)^2}{\La^4\over
Q^4}{\bar{I}}_2^{pv} \label{noz}.
\end{eqnarray}
We would like to emphasize that we could have chosen some other prescription for defining the
integral over $\sigma$ and hence the overall scale of the power corrections is arbitrary.  An
important aspect of Eq. (\ref{noz}) is that the $z$-dependence of the power corrections differs
from the corresponding $z$-dependence of the power corrections due to
$\la\neq 0$ at the one-loop level (see Eqs. (\ref{ls}), (\ref{lf})). 
Thus, the predictions for the $z$-dependence of the power corrections are  sensitive to
the exact argument of the running coupling .

Let us focus first on the question concerning the most sensitive point of the derivation.
Imagine that the true argument of the running coupling is
in fact $\kp^2/(1-z)$, i.e,:
\be
\ap~\ra~\apz~.\label{change}
\ee
This would correspond to a change in $\La$ in the Eq. (\ref{borel}):
\be
\La^2~\ra~(1-z)\cdot \La^2
\ee
which clearly would bring Eq.(\ref{noz}) in line with the predictions based
on $\la^2\neq 0$, see Eqs. (\ref{ls}), (\ref{lf}).  The corresponding
principal value integral will be denoted by $I_n^{pv}$ and for future reference it is given by:
\begin{equation}
I_n^{pv}= (-1)^n{{\pi}^{3/2} \over \beta_1}\int_0^{s}{d\sigma \over
sin\pi\sigma}\left[{(4z{{\La}^2 \over Q^2})^{\sigma}(n+\sigma) \over
{\Gamma(1+n+\sigma)\Gamma(3/2-n-\sigma)}}-
{(4z{{\La}^2 \over Q^2})^{-\sigma}(n-\sigma) \over
{\Gamma(1+n-\sigma)\Gamma(3/2-n+\sigma)}}\right] \label{i}
\end{equation}

Since $F_L$ is not logarithmically enhanced  in the leading order, one may
think that there is no reason, a priori, to believe that the coupling in Eq. (\ref{renorm})
should run like $\ap$. However, since it is well known \cite{catani} that for the 
structure function $F_2$, perturbation theory upto the two loop order for the
leading contributions which are logarithmically enhanced as $z \rightarrow 1$  is consistent
with the identification $\ap$, let us consider the
$1/Q^2$ corrections to it. Here we will find not only the pattern indicated above but in addition
a different kind of arbritariness related to the fact that the structure function $F_2$  recieves
contributions which are  collinear divergent.

Using  dimensional regularization to isolate the collinear divergences we readily obtain 
the following expression for  the real contribution to $F_2$:
\begin{eqnarray}
F_2^{real}~=~2C_F{\as \over 2\pi}\int_0^{{Q^2(1-z) \over 4z}}{d{\kp}^2 \over {\kp}^2}
{\kp}^{-2\epsilon}{1 \over \sqrt{(1-z)^2-4(\kp^2/Q^2)z(1-z)}} \nonumber \\
- C_F{\as \over 2\pi}(1+z)\int_0^{{Q^2(1-z) \over 4z}}{d{\kp}^2 \over {\kp}^2}
{\kp}^{-2\epsilon}{{1-z} \over \sqrt{(1-z)^2-4(\kp^2/Q^2)z(1-z)}} \nonumber \\
-C_F{\as \over 2\pi}{3z \over {1-z}}{1 \over Q^2}\int_0^{{Q^2(1-z) \over 4z}}d{\kp}^2 
{1 \over \sqrt{(1-z)^2-4(\kp^2/Q^2)z(1-z)}} \nonumber \\
+C_F{\as \over 2\pi}(6z+4z^2){1 \over Q^2}\int_0^{{Q^2(1-z) \over 4z}}d{\kp}^2 
{1 \over \sqrt{(1-z)^2-4(\kp^2/Q^2)z(1-z)}} \label{f2real} .
\end{eqnarray}
There are of course the virtual contributions which in particular tranform factors 
like ${1 \over 1-z}$ into "$+$ " distributions, however, we omit them for the moment as they
are not crucial to the argument.  It is easy to check that integration over $\kp$ reproduces the
well known result for this structure function to leading twist. Note that 
only the integral multipying the first two terms
produces a collinear divergence proportional to $1/\epsilon$ ( $\epsilon$ is related to the
number of space time dimensions ($D$) by $D=4-2\epsilon$)  whereas the second type of
integral does not.

In order to determine the ${\La }^2/Q^2$ contributions from the above using the renormalon
method, let us first consider the result when we take $\as$ to be $\apz$ for all the terms on
the right hand side of Eq.(\ref{f2real}). Proceeding as for the case of $F_L$ we find:
\begin{equation}
F_2^{real} ={C_F \over 2\pi}\left[\left({2 \over 1-z} -
(1+z)\right) \left({Q^2 \over 4}{1-z \over z}\right)^{-2\epsilon}J 
 + \left(-{3 \over 4}{1 \over 1-z}+ ({3 \over 2}+z)\right)I\right], \label{fpower}
\end{equation}
where $J$ and $I$ are two different types of integrals given by:
\begin{eqnarray}
J=\int_0^{\infty}d\si\left({4z\La^2\over Q^2}\right)^{\si \beta_1}
{\Gamma(-\si \beta_1-\epsilon)\Gamma (1/2)\over \Gamma(1/2-\si \beta_1-\epsilon)} \\
I=\int_0^{\infty}d\si\left({4z\La^2\over Q^2}\right)^{\si \beta_1}
{\Gamma(1-\si \beta_1)\Gamma (1/2)\over \Gamma(3/2-\si \beta_1)}.
\end{eqnarray}
Notice the collinear divergence in $J$ for $\sigma \sim 0$ which was identified with the
perturbative region.  One may imagine defining $J$ and $I$ by two different prescriptions for
obtaining the power corrections ( for which the ratios $J/I$ would be different )
and in this way we see that two arbitrary scales would appear
in the predictions.  Thus if $J_n$ and $I_n$ denote the appropriately defined integrals in the
integration region over $\sigma$ which produces the power correction $(1/Q^2)^n$, then:
\begin{equation}
\left[F_2^{real}\right]_{1/Q^2} ={C_F \over 2\pi}\left[\left({2 \over 1-z} -
(1+z)\right) 4zJ_1 
 + \left(-{3 \over 4}{1 \over 1-z}+ ({3 \over 2}+z)\right)4zI_1\right]
{{\La}^2 \over Q^2}. \label{fpower2}
\end{equation}
In the principal value prescription, for example,
  $J_n^{pv}$ is different from $I_n^{pv}$ (see Eq.(\ref{i}) ) and is given by : 
\begin{equation}
J_n^{pv}~=~(-1)^n{\pi^{3/2} \over \beta_1}\int_0^{s}{d\sigma \over
sin\pi\sigma}\left[{(4z{{\La}^2 \over Q^2})^{\sigma}(n-1/2+\sigma) \over
{\Gamma(1+n+\sigma)\Gamma(3/2-n-\sigma)}}-
{(4z{{\La}^2 \over Q^2})^{-\sigma}(n-1/2-\sigma) \over
{\Gamma(1+n-\sigma)\Gamma(3/2-n+\sigma)}}\right]  \label{j}
\end{equation}
If we would have used the running $\ap$ instead of $\apz$ then in particular, 
we would have an extra factor of $1-z$ in the denominators of Eq. (\ref{fpower2}).
The $z$ dependance of the higher twist contributions as given in Eq. (\ref{fpower2}) is of a
similar type as that obtained from the method of non zero gluon mass \cite{dasgupta}, however
in this case we find that the expression for the power corrections involves two arbritary scales
$I(\La^2/Q^2)$ and $J(\La^2/Q^2)$ rather than a single gluon mass parameter $\la$. 
The predictions for the $z$  dependance will be different depending on which prescription
we choose. The origin of these two different functions is that $J$ arises from integrals which
are collinear divergent and $I$ from those that do not have this divergence.  It is therefore not
surprising to expect different functions appearing in this manner. This situation is not
improved if we let the coupling in the various terms of Eq.(\ref{f2real}) run in a different
manner depending on whether they come from collinear or soft regions as $z \rightarrow  1$.

There are two aspects of the above discussion that we would like to emphasize.
 (1) The predictions for the observable
$z$ dependence of  the power corrections depend crucially 
on the argument of the running coupling. For the case of $F_L$, if the coupling runs
as $\ap$ then the results disagree with the case $\la^2\neq 0$ 
while the running as $\apz$  reproduces the results of the previous
section ( see also \cite{dmw}). The argument of the running coupling can be clarified in
perturbation theory  by two-loop calculations and has been studied for say the structure
function $F_2$ in a number of papers as mentioned above,\cite{catani}.
 (2) More importantly, we see that for the structure function $F_2$
the renormalon technique necessitates the introduction of two unknown  scales(associated
respectively with $I$ and $J$) rather than just one in the case of $\la^2 \neq  0$.
Thus keeping in mind the transition to the non-perturbative case we should reserve for at least
two independant rescaling functions. In this respect the picture is different from that in
section 2. We will see a furthur indication of this in the next sections.

\section{Quark Mass as an Infrared Parameter.}

In an attempt to further quantify the infrared sensitivity in deep inelastic scattering
phenomena, in this section, we consider the quark mass as an infrared parameter
substituting  say the gluon mass of section 2.
It is convenient to discuss the role of the quark mass 
for the moments of the structure functions,
i.e. in the language of OPE. Moreover in this language, as discussed at the end of this section, it
is most convenient to find the  possible sources of infrared instability.
 In order to make the connection to earlier work
\cite{mueller1} which also exploited the OPE more transparent we choose to consider in this
section, the  moments of the antisymmetric structure function $F_3$.

The forward Compton amplitude
has the standard operator product expansion:
\bea
T_{\mu\nu}~=~i\int d^4xexp(iqx)T(j_{\mu}(x)j_{\nu}^{\dagger}(0))~=~
\sum_{i,n}\left({2\over -q^2}\right)^n\cdot
(g_{\mu\nu}-{q_{\mu}q_{\nu}\over q^2})q_{\mu _1}q_{\mu_ 2}C^i_L-\\ \nonumber
-(g_{\mu \mu_1}g_{\nu\mu_2}q^2-g_{\mu\mu_1}q_{\nu}q_{\mu_2}-
g_{\nu \mu_2}q_{\mu}q_{\mu_1}+g_{\mu\nu}q_{\mu_1}q_{\mu_2})C^i_2
-\eps_{\mu\nu\mu_1\alpha}q_{\alpha}q_{\mu_2}C^i_3)
q_{\mu_3}...q_{\mu_n}\theta ^i_{\mu_1...\mu_n}.\label{sum}
\eea
where $C_i$ are the
coefficient functions and the $\theta$ are the operators.
We will be interested in the $1/Q^2$ correction keeping the 
leading twist contribution as well.
The calculation of section 2
which reproduces the results of Ref. \cite{dmw} corresponds to a two-step
procedure in evaluating the matrix elements of the operators $\theta$ \cite{stein}.
Namely, one first reduces the operators of higher twist to those
of the leading twist and then expresses the latter in terms of
phenomelogical structure functions in a standard way.
  
We will explain this procedure in detail on the  
example of the lowest moment of $F_3$.
The relevant operator is \cite{vs,soldate}:
\be
T^A_{\mu\nu} 
~=~{2i\over q^2}\eps_{\mu\nu\alpha\beta}q_{\alpha}\left(L_{\beta}+{4\over q^2}\theta_{\beta}
\right)
\ee
where
\begin{eqnarray}
L_{\beta}~=~\bar{q}\gamma_{\beta}q,  \label{lb}\\
\theta_{\alpha}~=~g_s\bar{q}\tilde{G}^a_{\alpha\beta}t^a\gamma_{\beta}\gamma_5q, 
\label{tb} \\
\tilde{G}^a_{\alpha\beta}~=~{1\over 2}\eps_{\alpha\beta\gamma\de}G^a_{\gamma\de}.
\end{eqnarray}
Now, we assume that the matrix element of the operator $\theta_{\beta}$ containing
the gluon field can be reduced to that of $L_{\beta}$ by using perturbation theory:
i.e, we simply evaluate the contributions from the diagrams of Fig.1
with quark external states. (The open
circles represent the insertion of the higher twist operator). In this way we get,
\be
\langle p|T^A_{\mu\nu}| p\rangle~\approx~
{2i\over q^2}\eps_{\mu\nu\alpha\beta}q_{\alpha}
\left(\langle p|L_{\beta}|p\rangle - f(\eps^2,m^2\la^2){C_F\over2\pi}
{4\as\over3}\langle p|L_{\beta}|p\rangle \right)\label{ope}
\ee
where the factor $f(\eps^2,m^2,\la^2)$ already mentioned in the Introduction is:
\be
f(\eps^2,m^2,\la^2)~=~{2\over q^2}\int_0^1dx M^2ln(M^2),~~~
M^2~=~-\eps^2x(1-x)+m^2x^2+\la^2(1-x)\label{factor}.
\ee
The evaluation of 
the $1/Q^2$ correction considered in section 2 \cite{dmw,stein,dasgupta} corresponds 
to $f(0,0,\la^2)=-2(\la^2/Q^2)ln\la^2$.

Eq. (\ref{factor}) demonstrates an origin of the renormalon ambiguities in DIS.
From the point of view of the general OPE (see  Eq. (\ref{sum}))
there is no reason whatsoever
to treat the operators of the quark and gluon
fields in an asymmetric way. The assumption that one can first integrate
out the gluon line perturbatively and use a non-perturbative input on
the matrix element of quark fields is arbitrary. This can be contrasted
with the basic idea of renormalons\cite{mueller,vz} when 
only one gluon line is made soft by means of an insertion of a large
number of vacuum bubbles. Now we have both gluon and quark lines soft.
It is only in this way that one can get $1/Q^2$ corrections.
Thus the estimates of the $1/Q^2$ corrections in DIS can be understood
only within a more general framework when one introduces infrared
sensitive parameters in various ways. Then the quark mass is no worse
a parameter than the gluon mass and one can look for $m^2lnm^2$ terms 
instead of $\la^2ln\la^2$ terms. However for the case of the quark mass
must be careful to take into account the purely kinematic higher twist
contributions a la Nachtman\cite{nachtman}. To illustrate this point let us
consider the second moment of the structure function $F_3$.

The relevant operator now is \cite{vs,soldate}:
\be
T^{A}_{\mu\nu}= {4i \over q^4}\epsilon_{\mu\nu\alpha\beta}q_{\alpha}q_{\gamma}
(L_{\beta\gamma} - {3 \over {32q^2}}A_{\beta\gamma} + {7 \over {16q^2}}B_{\beta\gamma})
\ee
where, the leading twist operator is also included and symmetrization and trace
subtractions are not explicitly shown. Furthur, 
\bea
\L_{\beta\gamma}= i\bar{q}\gamma_{\beta}D_{\gamma}q \\
A_{\beta\gamma}= -2g \bar{q}D_{\alpha}G_{\alpha\gamma}^at^a \gamma_{\beta}q \\
B_{\beta\gamma}= g \bar{q}\{\tilde{G}_{\beta\alpha}^a, iD_{\gamma}\}_{+}t^a
\gamma_{\alpha}\gamma_{5}q  
\eea
Just as before, the matrix elements of
the operators containing the gluon field can be reduced to that of $L_{\beta\gamma}$
by evaluating the contributions from Fig.1. The results for the two operators are:
\bea
\langle A_{\beta\gamma}\rangle= {C_F \over {2\pi}}\alpha_s 
{4i\epsilon_{\mu\nu\alpha\beta} q_{\alpha}q_{\gamma} \over q^4}
\left( {{f_1^a(p^2,m^2,\la^2) +
f_2^a(p^2,m^2,\la^2)} \over q^2}\right)\langle L_{\beta\gamma}\rangle \label{ma}\\
\langle B_{\beta\gamma}\rangle = {C_F \over {2\pi}}\alpha_s { f^b(p^2,m^2,\la^2) \over
q^2}\langle L_{\beta\gamma}\rangle 
\eea
where,
\bea
f_1^a(p^2,m^2,\la^2) = 9\int_0^1 dx (1-x) M^2 ln M^2 \\
f_2^a(p^2,m^2,\la^2) = -3\int_0^1 dx x (-p^2x(1-x)+m^2x) ln M^2 \label{f2}\\
f^b(p^2,m^2,\la^2) = -28\int_0^1 dx x M^2 ln M^2
\eea
It is easy to verify that for $p^2=m^2=0$ the results of ref.\cite{dasgupta} are
reproduced. For finite quark masses,  we see 
from Eq.(\ref{ma}) and Eq,(\ref{f2}), that it appears in a 
different way than the gluon mass through the function $f_2^a$. The origin of 
this function 
however is easily understood as a consequence of a kinematic effect due to finite
quark masses. To see this we first note that in general the coefficient function 
$C(q)$ will depend on the running quark mass,
\be
m(Q)= m(Q_0) + m(Q_0)\gamma_mln{Q \over Q_0}
\ee
where, $\gamma_m$ is the mass anomalous dimension which at the one loop order is:
\be
\gamma_m=-3{C_F \over {2\pi}}\alpha_s 
\ee
The coefficient function itself may be expanded:
\be
C(q)=C^{0}(q,m) + \gamma_m ln{Q \over m}m{\partial \over {\partial m}}C^0 + ....
\ee
where the ellipses denote other higher order terms. The mass dependance of the 
coefficient functions can be inferred to any order in $m/Q$ 
 by considering the structure functions as a
function of the  Nachtman variable $\xi$: \cite{nachtman,georgi}
\be
\xi = x[1/2 + (1/4+m^2/Q^2)^{1/2}]
\ee
Using this it is straightforward to reproduce the contribution $f_2^a$.
Thus we see that when the purely kinematic effects due to the quark masses are taken into
account following a procedure analogous to that of  Ref. \cite{nachtman}, we get similar
predictions for the infrared sensitivity keeping $m$ as an infrared parameter or
$\la^2 \neq 0$, at least upto the second moment of $F_3$.

 The formulation discussed in this section is  the appropriate language for the inclusion
of the anomalous dimensions of the operators governing the power corrections. A 
discussion of this,  as we see next,brings into question the infrared safety of the 
relations like those in Eqs. (\ref{ls}), (\ref{lf}) and (\ref{ope}).

Let us go back to the case of the first moment of $F_3$. The leading 
twist contribution and 
the first power corrections are governed then by the operators 
$L_{\beta}$ and $\theta_{\beta}$, respectively (see Eqs. (\ref{lb}, \ref{tb})).
The anomalous dimensions of the operator $L_{\beta}$ is obviously vanishing
while that  of the operator $\theta_{\beta}$ is \cite{vs}:
\be
\gamma_{\theta}~=~ -{\as \over 4\pi}({ 32 \over 9})
.\ee
Since the anomalous dimensions of $L_{\beta}$,  and $\theta_{\beta}$
are different, Eq. (\ref{ope}) can at best be true only for
a particular choice of $Q^2$. In other words, the relations like Eqs. (\ref{ope}) and
in general, (\ref{ls}), (\ref{lf}) and
 are not infrared safe. If one accounts first for a splitting of the quark 
into a collinear quark and gluon and 
then for the emission of a soft gluon, 
the corresponding contribution to the
power correction is not suppressed by $\aQ$.
This is the meaning of a nonvanishing anomalous dimension of
$\theta_{\beta}$.
Note that in this respect the properties of the power like corrections
differ from the properties of the leading twist contributions. In the latter case 
there are no new collinear divergences in higher loops if one considers 
the non-singlet DIS structure functions.
This  in turn means that the power corrections are 
governed by an independent structure function  and its reduction  to the leading twist one
cannot be justified in any known approximation  ( for an earlier discussion, see \cite{ellis}). As
mentioned in section 3,  an indication of this is already present in the renormalon chain method
of obtaining the power corrections where  these were found to  involve two different scales 
through the integrals $I$ and $J$.

\section{Conclusions.}
The picture of predictions for the power-like corrections to
deep inelastic scattering is remarkably rich.
From a  purely  phenomenological viewpoint, one deals not with a few 
numbers fixed by renormalons as is usually the case but
rather with a few functions which are x-dependences of the
power corrections  to various structure functions \cite{dmw,stein, dasgupta}.
This makes comparison with the experimental data
much more challenging. On the theoretical side, as is emphasized in this paper,
there is an unusual variety of possible choices of infrared sensitive parameters. 
We considered in detail the consequences of choosing the gluon and quark masses
as well as the position of the Landau pole in the running coupling as infrared
parameters. To summarize the results, let us mention some of the conclusions made:

(1) Keeping the gluon mass $\la\neq 0$ and defining the infrared sensitive
contributions as those non-analytic in $\la^2$,  results exactly in the
same predictions as the dispersive approach to the coupling \cite{dmw} and 
the renormalon chain in the large $N_f$ limit \cite{stein}. The choice between these
techniques is purely a matter of convenience and the case of $\la\neq 0$ looks
most simple from the computational point of view.

(2)  Parametrizing the power correction in terms of $\La$ brings out two interesting features.
The predictions  based on $\la\neq 0$ are reproduced if one assumes
that the coupling runs as $\as (k_{\perp}^2/(1-z))$.
On the other hand, only by letting the coupling  run as $\ap$ do we get consistency with explicit
two-loop calculations  in case of the structure function $F_2$.  . This latter choice results in a
different pattern of the power corrections to $F_2$ and to $F_L$ than the case $\la \neq 0$.
The power correcttions to $F_L(x,Q^2)$ are more important phenomenologically
since $F_L(x,Q^2)$ is proportional to $\aQ$ in the leading-twist approximation.  In fact 
here at the two loop level,  one would encounter diagrams of the type shown in Fig. 2
which can be shown to have power corrections of the type ${\la^2 \over {Q^2(1-z)}}$ at the
partonic level. It is not clear that such contributions can be neglected even though they arise at
a higher perturbative order. We also found that  because of the collinear divergence , not
one but two unknown scales are ,  in general, introduced in the renormalon chain method. 
One which is associated with $I_1 \cdot {{\La}^2 \over Q^2}$ and the other with
$J_1 \cdot {{\La}^2 \over Q^2}$. This again is different from the case $\la \neq 0$
and is indicative of an inherent infrared instability.

(3) The use of the quark mass to identify the infrared sensitive contribution
results, generally speaking, in a different pattern of
power-like corrections.
The use of the hypothesis on non-perturbative enhancement of 
the infrared sensitive contributions \cite{az2} is crucial at 
this point. Namely one splits $m^2ln~m$ terms into two pieces.
One piece is a power-like corrections due to a finite mass of
the target- the quark in our case. Such contributions may be treated in a manner similar to that
in \cite{nachtman}. The other piece is to be treated as a signal
for an infrared sensitive contribution associated with low $k_{\perp}^2$.
Allowing for a non-perturbative enhancement of this
piece brings the predictions in line with the
case $\la\neq 0,m=0$.

(4) From the point of view of the OPE approach, the calculations of sections 2, and 3  
correspond to an asymmetrical treament of the  effects
due to the soft gluon and quark lines.
One integrates over the gluon line perturbatively while
the quark distribution (as manifested in the structure functions) are borrowed from
experiment which implies the inclusion of both perturbative and non-perturbative
infrared effects. The dependence on the quark virtuality inherent to the
reduction factor ({\ref{factor}) signals that the procedure cannot in fact be
substantiated theoretically. However, the virtuality of the quark is not a
convenient parameter because it is manifestly not gauge invariant.
To circumvent this we discussed the problem in the framework of the anomalous dimension of
the operators governing the power corrections. Since these are different for the leading twist
and the higher twist operators,  we concluded that in general we are dealing with new infrared
unsafe quantities at the level of the power corrections which cannot be cured by just
introducing the same  structure function as for the leading twist.  Hence independant structure
functions are required for the power corrections. Thus, experimental confirmation 
of the one-loop results given in \cite{stein,dasgupta}  and  reviewed in section 2 
of the paper  would in fact bring about a puzzle because of 
a successful perturbative evaluation of an infrared unsafe quantity.
One would then be confronted with the challenge of 
formulating the approximation involved in more precise terms.

\section{Acknowledgements} 
We would like to thank S. Catani for an interesting discussion. We would also like to thank
Yu. Dokshitzer and G. Marchesini for communications concerning  Ref. \cite{dmw}.
This work was supported in part by the US Department of Energy.

\newpage 

\centerline{ 
\epsfxsize=0.6\textwidth 
\epsffile{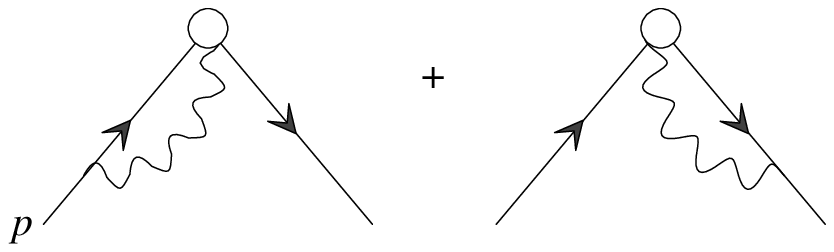} 
}
\begin{center} 
{\large \bf Fig.~1} 
\end{center} 

\vspace{1.0cm}
 
\centerline{ 
\epsfxsize=0.6\textwidth 
\epsffile{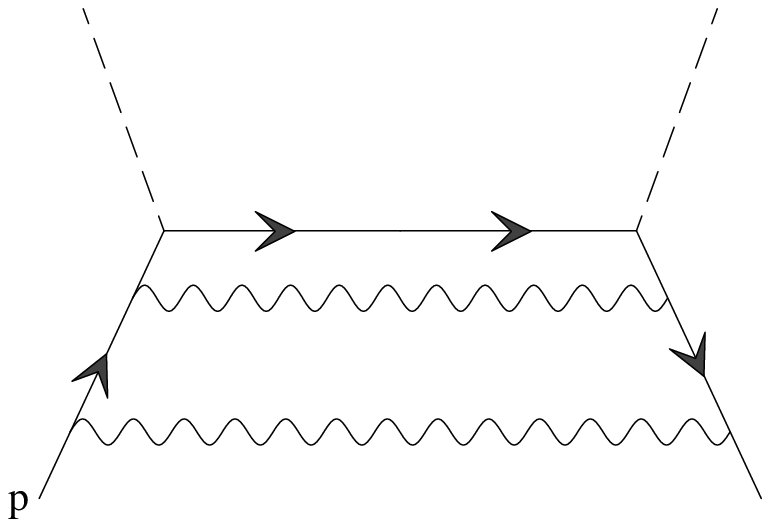} 
}
\begin{center} 
{\large \bf Fig.~2} 
\end{center}


\begin{thebibliography}{99}
\bibitem{footnote} The results of this paper were briefly summarized in Ref. [3].
\end{thebibliography}
\end{document}